\RequirePackage[2020-02-02]{latexrelease}
\documentclass[amssymb,amsfonts,amsmath,reprint,aps,prapplied, superscriptaddress]{revtex4-2}
\usepackage{graphicx} 

\usepackage{amsmath} 
\allowdisplaybreaks
\usepackage{amsmath}
\usepackage{physics}
\usepackage{amsfonts}
\usepackage{subfigure}
\usepackage{MnSymbol}
\usepackage{xcolor}
\usepackage{comment}  
\usepackage{hyperref}
\usepackage{wasysym}
\usepackage{appendix}
\DeclareMathOperator{\sgn}{sgn}
\usepackage{url}

\begin{document}
    
    \title{\bf Fast high-fidelity charge readout by operating the cavity-embedded Cooper pair transistor in the Kerr bistable regime}
    \date{\today}
    
    \author{B. Thyagarajan}    
    \email{Bhargava.Thyagarajan.gr@dartmouth.edu}
    \affiliation{Department of Physics and Astronomy, Dartmouth College, Hanover, New Hampshire 03755, USA}
    \author{S. Kanhirathingal}
    \affiliation{Department of Physics and Astronomy, Dartmouth College, Hanover, New Hampshire 03755, USA}
    \affiliation {Rigetti Computing, Berkeley, California 94710, USA}
    \author{B.L. Brock}
    \affiliation{Department of Physics and Astronomy, Dartmouth College, Hanover, New Hampshire 03755, USA}
    \affiliation{Department of Applied Physics, Yale University, New Haven, Connecticut, USA}
    \author{Juliang Li}
    \affiliation{Department of Physics and Astronomy, Dartmouth College, Hanover, New Hampshire 03755, USA}
    \affiliation {High Energy Physics Divison, Argonne National Laboratory, 9700 South Cass Avenue, Argonne, IL 60439, USA}
    \author{M.P. Blencowe}
    \affiliation{Department of Physics and Astronomy, Dartmouth College, Hanover, New Hampshire 03755, USA}
    \author{A.J. Rimberg}
    \email{Alexander.J.Rimberg@dartmouth.edu}
    \affiliation{Department of Physics and Astronomy, Dartmouth College, Hanover, New Hampshire 03755, USA}

    \begin{abstract}
    Operating the cavity-embedded Cooper pair transistor (cCPT) in the Kerr bistable regime, we demonstrate single-shot resolution between two charge states that are $0.09e$ apart. The measurement is performed with 94$\%$ fidelity in a duration of 3 $\mu$s. The drive power at which the measurement is performed corresponds to only 20 intracavity photons on average in the high oscillation amplitude state of the cCPT, which is orders-of-magnitude smaller than that in rf-SETs. We find that the limiting factor for this mode of operation of the cCPT is the spontaneous fluctuation-induced switching between the two metastable oscillation amplitude states. We present empirical data on the variation of the switching dynamics with drive parameters and cCPT DC bias.
    \end{abstract}
    
    \maketitle
    
    \section{Introduction}
    \label{introduction_section}
    Fast detection of charge on the order of a fraction of an electron has long been an important task. Versatile devices such as the quantum point contact and the single electron transistor (SET) have been used to measure electron lifetimes in a single electron trap \cite{Dresselhaus_PRL_1994}, to map electric fields with 100 nm spatial resolution \cite{Yoo_science_1997}, to observe macroscopic charge quantization \cite{Lafarge_Z_physik_B_CM_1991}, and to study quasiparticle and electron tunneling events in real-time \cite{Naaman_PRB_2006, Lu_Nature_2003}. More recently, they have been used in the search for Majorana zero modes in nanowires \cite{vanZanten_Nature_phys_2020}, and could potentially be used to detect dark matter \cite{Barak_PRL_2020, essig_arxiv_2021}. Such fast, ultrasensitive electrometers are instrumental in the readout of silicon-based spin qubits \cite{Morello_Nature_2010, He_nature_2019} where the magnetic moment of a single spin is too small to detect directly, and is instead converted to a spin state dependent charge which can be read out. Dispersive charge sensors operating on the supercurrent branch of the Josephson junctions based inductive-SET (L-SET) \cite{Sillanpaa_PRL_2004} and the single Cooper-pair box \cite{Persson_PRB_2010} are not shackled by the electron shot noise which limits the operation of the rf-SETs \cite{Schoelkopf_Science_1998}. The cavity-embedded Cooper pair transistor (cCPT) used in this work has previously been shown to achieve a charge sensitivity of 14 $\mu e/\sqrt{\mathrm{Hz}}$ operating as a dispersive sensor in the linear regime with a single intracavity photon on average \cite{Brock_PhysRevApplied_2021_ultrasensitive}, close to the theoretical quantum limit for this device \cite{kanhirathingal_JAP_2021}. 
    
    \par
    The cCPT is also a rich nonlinear system whose Hamiltonian contains a Kerr nonlinearity \cite{Brock_PhysRevApplied_2021}, and an emergent parametric amplifier term when the flux line of the system is driven at twice the resonance frequency.
    The Kerr term opens up the possibility of more sensitive charge detection than was achieved in the linear regime \cite{Tosi_Phys_rev_applied_2019}. Such a Kerr cavity coupled to a mechanical resonator was proposed \cite{Nation_PRB_2008} and demonstrated \cite{Zoepfl_Arxiv_2022} to achieve an order of magnitude better cooling of the phonon mode compared to a linear cavity. The Kerr nonlinearity is well known to produce bifurcations in the system response \cite{Nayfeh_book_1995}. Bifurcation amplifiers \cite{Siddiqi_PRL_2005, Siddiqi_PRL_2004, Vijay_Sci_rev_instrum_2009} based on a large change in the response at a bifurcation edge have been used to read out the state of superconducting qubits \cite{Mallet_Nat_Phys_2009}. Nanomechanical devices based on the bifurcation under a parametric drive have been used to sense charges of $~9e$ at room temperature \cite{DashAPL2021}. Similar devices have demonstrated charge sensing of the order of $70e$ by manipulating the topology of the bifurcation diagram \cite{Karabalin_PRL_2011}. 
    
    \par
    Here, using the bifurcation between a bistable and a monostable region induced by the Kerr nonlinearity of the cCPT, we demonstrate single-shot readout of $0.09e$ of charge in 3 $\mu$s with 94$\%$ fidelity, using fewer than 25 intracavity photons. Such low power operation ensures minimal back-action on the system being measured \cite{Aassime_PRL_2001}, and also aids in the integration of these cCPT detectors with state-of-the-art first stage amplifiers such as the TWPAs \cite{Macklin_science_2015} without overwhelming them beyond their compression point. Such fast high fidelity readout is comparable to the current state-of-the-art for semiconductor spin qubits \cite{keith_PRX_2019, DAnjou_PRB_2019}.
    
    \par
    In Sec. \ref{theory_section} we present a semi-classical analysis of the nonlinear cCPT and propose a scheme for it to function as a sensitive charge state discriminator operating in the bistable regime. In Sec. \ref{experiment_section} we experimentally study the hysteresis in the cCPT response in the bistable regime to characterize the extent of the bistability as a function of the drive detuning and strength. We then implement a charge sensing protocol, and observe the presence of fluctuation-induced spontaneous transitions between the bistable states, which we study as a function of drive parameters and cCPT DC bias. Lastly, we characterize our charge sensing protocol and demonstrate the optimum high-fidelity, fast charge state readout possible with this device. In Sec. \ref{discussion_section} we conclude by discussing some possible improvements to this work. Details of the heterodyne measurement scheme employed in this work and the microwave circuitry used in the dilution refrigerator are in Appendix \ref{appendix_experiment_circuit}, and some experimental considerations for the charge sensing scheme used in Sec. \ref{experiment_section} are detailed in Appendix \ref{appendix_charge_sensing_protocol}.
    
    \begin{figure}[h!]
        \centering
        \includegraphics[width = 0.5\textwidth]{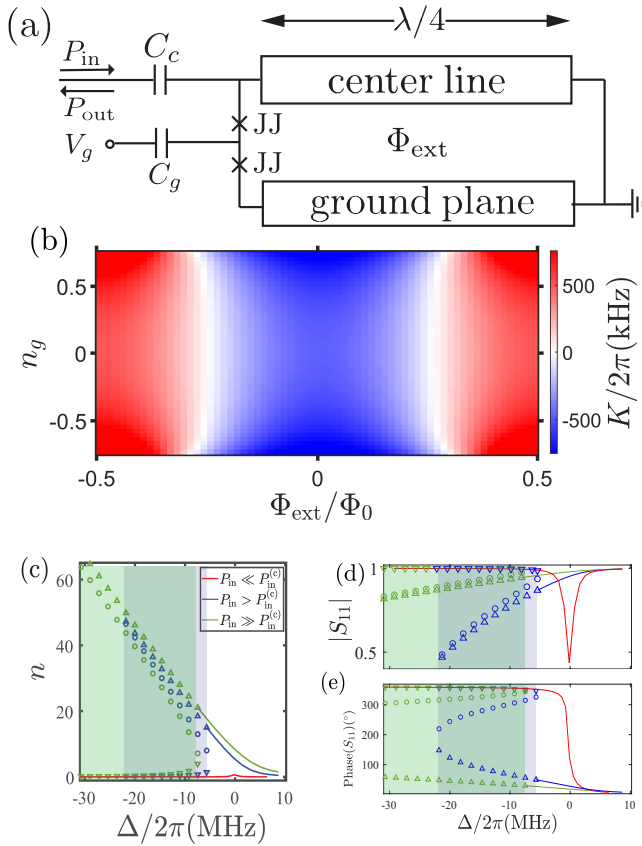}
        \caption{(a) Schematic of the cCPT. (b) Variation of the Kerr coefficient $K$ as a function of gate, $n_g$, and flux, $\Phi_{\mathrm{ext}}$, over the operational bias range of the cCPT simulated using the extracted values of $E_J$ and $E_C$ \cite{Brock_PhysRevApplied_2021}. (c) Simulated response of the cCPT for different drive powers, $P_{\mathrm{in}}$. The red curve is for a drive strength $P_{\mathrm{in}} \ll P_{\mathrm{in}}^{\mathrm{(c)}}$. The blue curve is for $P_{\mathrm{in}} > P_{\mathrm{in}}^{\mathrm{(c)}}$ and the green is for $P_{\mathrm{in}} \gg P_{\mathrm{in}}^{\mathrm{(c)}}$. Above $P_{\mathrm{in}}^{\mathrm{(c)}}$, we see bistability across a range of detunings indicated by the corresponding shaded region. The $\bigtriangleup$'s represent the stable high oscillation amplitude state, the $\bigtriangledown$'s represent the stable low oscillation amplitude states, and the $\Circle$'s represent the unstable state. The solid lines indicate monostability. (d) Simulated magnitude of the reflection coefficient for the drive powers in (c). (e) Simulated phase of the reflection coefficient for the drive powers in (c). All simulations in (c), (d) and (e) were for a cCPT DC bias $(n_g, \Phi_{\mathrm{ext}}) = (0, 0)$ with $K/2\pi = -470$ kHz, and nominal damping rates for this bias point \cite{Brock_PhysRevApplied_2021}.}
        \label{fig_1_composite}
    \end{figure}
    
    \section{Theoretical Description}
    \label{theory_section}
    The cCPT, schematically depicted in Fig. \ref{fig_1_composite}(a), consists of two parts: (i) the cavity, which is a $\lambda/4$ superconducting coplanar waveguide (CPW) with its end shorted to the ground plane, and (ii) a Cooper pair transistor (CPT) across the center line and ground plane of the CPW at its voltage anti-node. In this geometry, the cavity and CPT form a shared SQUID loop, which couples them together. When the CPT remains in its ground state, it modifies the effective potential of the cavity, such that the cCPT behaves as a nonlinear oscillator whose resonant frequency can be tuned using the effective gate charge, $n_g = \frac{C_g V_g}{e}$, and the magnetic flux threading the SQUID loop, $\Phi_{\mathrm{ext}}$. Here, $V_g$ is the external DC voltage applied to the CPT island through the gate capacitance $C_g$. Along with the fabrication details for the cCPT device used in this work, a detailed characterization at low drive amplitudes where the nonlinearities do not contribute substantially to the dynamics has been carried out in Ref. \cite{Brock_PhysRevApplied_2021}. Notably, the Josephson energy, $E_J$, and the charging energy, $E_C$, were estimated to be $E_J/h = 14.8$ GHz and $E_C/h = 54.1$ GHz respectively. Finally, to drive and measure the cCPT, a probe transmission line is coupled to the CPW through a coupling capacitor $C_c$.
    
    \par 
    For an input drive close to resonance, under the rotating wave approximation, the Hamiltonian for the cCPT is given by \cite{Brock_PhysRevApplied_2021, kanhirathingal_JAP_2021} 
    \begin{align}
       H &= \hbar\omega_0(n_g, \Phi_{\mathrm{ext}})a^\dagger a + \frac{1}{2} \hbar K(n_g, \Phi_{\mathrm{ext}}) a^{\dagger2} a^2,
    \label{kerr_cavity_hamiltonian}
   \end{align}
   where $a(a^\dagger)$ are the annihilation (creation) operators for the cavity mode, $\omega_0(n_g, \Phi_{\mathrm{ext}})$ and $K(n_g, \Phi_{\mathrm{ext}})$ are the resonant frequency of the linear cCPT Hamiltonian and the Kerr coefficient respectively. The variation of $K(n_g, \Phi_{\mathrm{ext}})$ over the operational range of the cCPT device used in this work is shown in Fig. \ref{fig_1_composite}(b). The Kerr coefficient changes sign with flux, attaining extremum values at half-integer multiples of $\Phi_0$, and passing through zero close to $\Phi_{\mathrm{ext}} = 0.25\Phi_0$. Kerr-free cavities have been used to increase the dynamic range of parametric amplifiers \cite{Frattini_Phys_rev_app_2018,Sivak_Phys_Rev_App_2019}. 
   
   \par 
   We use input-output theory \cite{Gardiner_PRA_1985} to model the dynamics of the cavity mode. The quantum Langevin equation for $a$ gives
   \begin{align}
      \dot{a} &= \frac{1}{i\hbar} \big[ a, H\big] - \big[ a, a^\dagger\big] \left(\frac{\kappa_{\mathrm{tot}}}{2} a - \sqrt{\kappa_{\mathrm{ext}}}a_{\mathrm{in}}(t) - \sqrt{\kappa_{\mathrm{int}}}b_{\mathrm{in}}(t) \right) \nonumber \\
      &= -\left(i\left(\omega_0 + K a^\dagger a\right) + \frac{\kappa_{\mathrm{tot}}}{2}\right)a  + \sqrt{\kappa_{\mathrm{ext}}} a_{\mathrm{in}}(t) + \sqrt{\kappa_{\mathrm{int}}}b_{\mathrm{in}}(t)  ,
      \label{kerr_hamiltonian_quantum_langevin_equation_a}
    \end{align}
    and a corresponding equation for $a^\dagger$, where $\kappa_{\mathrm{ext}}$ is the external damping rate due to the coupling of the resonator to the probe transmission line with the input bath operator $a_{\mathrm{in}}(t)$, and $\kappa_{\mathrm{int}}$ is the internal damping rate associated with the coupling of the resonator to an internal loss channel with input operator $b_{\mathrm{in}}(t)$. The total damping rate of the cavity is $\kappa_{\mathrm{tot}} = \kappa_{\mathrm{ext}} + \kappa_{\mathrm{int}}$. When the input tone is a pure sine wave at frequency $\omega_d$ of the form $\langle a_{\mathrm{in}} \rangle = \alpha_{\mathrm{in}}e^{-i\omega_dt}$, the steady state response of the cavity is at this drive frequency. For this coherent drive, using the semi-classical approximation, we make the ansatz $\langle a \rangle = \alpha e^{-i\omega_d t}$, with $\langle\dot{a} \rangle = -i\omega_d\alpha e^{-i\omega_dt}$ and the average intracavity occupation number $n = \abs{\alpha}^2 = \langle a^\dagger a \rangle$. Plugging this ansatz into the expectation value of Eq.(\ref{kerr_hamiltonian_quantum_langevin_equation_a}) we obtain 
  \begin{align}
     \left[-i\left(\Delta - K\abs{\alpha}^2\right) + \frac{\kappa_{\mathrm{tot}}}{2} \right] \alpha &= \sqrt{\kappa_{\mathrm{ext}}} \alpha_{\mathrm{in}},
     \label{kerr_hamiltonian_steady_state_solutions}
  \end{align}
    where we have defined the detuning $\Delta = \omega_d - \omega_0$. Using Eq.(\ref{kerr_hamiltonian_steady_state_solutions}) and the input-output relation $a_{\mathrm{out}}(t) = a_{\mathrm{in}}(t) - \sqrt{\kappa_{\mathrm{ext}}}a(t)$ \cite{Gardiner_PRA_1985}, \cite{text_gardiner_zoller} we find the reflection coefficient $S_{11}(\Delta)$ to be
  \begin{align}
       S_{11}(\Delta) &= \left(\frac{\alpha_{\mathrm{out}}}{\alpha_{\mathrm{in}}}\right) ^* \nonumber  \\
       & = \frac{(\Delta - K\abs{\alpha}^2) - i(\kappa_{\mathrm{int}} - \kappa_{\mathrm{ext}})/2}{(\Delta - K\abs{\alpha}^2) - i(\kappa_{\mathrm{int}} + \kappa_{\mathrm{ext}})/2},
       \label{kerr_cavity_S11}
  \end{align}
  where $a_{\mathrm{out}}(t)$ is the transmission output bath operator. Also, from Eq.(\ref{kerr_hamiltonian_steady_state_solutions}) and the corresponding equation for $\alpha^*$, $n = \abs{\alpha}^2$ satisfies the cubic equation
  \begin{align}
  K^2n^3 - 2K\Delta n^2 + \left(\Delta^2 + \frac{\kappa_{\mathrm{tot}}^2}{4} \right) n = \kappa_{\mathrm{ext}} \frac{P_{\mathrm{in}}}{\hbar \omega_d},
  \label{number_photons_cubic_equation}
  \end{align}
  where $P_{\mathrm{in}} = n_{\mathrm{in}}\hbar \omega_d$ is the power of the input drive tone incident on the cCPT, and $n_{\mathrm{in}} = \abs{\alpha_{\mathrm{in}}}^2$ is the input photon flux. As illustrated in Figs. \ref{fig_1_composite}(c-e), at very low drive strengths this cubic equation has only one real root and the oscillator exhibits only monostable behaviour across all detunings. As the drive strength is increased beyond a critical power $P_{\mathrm{in}}^{\mathrm{(c)}} = \frac{\sqrt{3}}{9} \frac{\kappa_{\mathrm{tot}}^3}{\abs{K}\kappa_{\mathrm{ext}}}\hbar\omega_{d}^{(c)}$, the oscillator system undergoes a bifurcation, and exhibits bistability for a range of detunings. Here, $\omega_{d}^{(c)}$ is the drive frequency corresponding to a detuning of $\Delta_c = \sgn({K})\frac{\sqrt{3}}{2}\kappa_{\mathrm{tot}}$ and $(\Delta_c, P_{\mathrm{in}}^{\mathrm{(c)}})$ corresponds to a spinode point in the parameter space of the input drive \cite{DykmanPhysica1980}. In the bistable region, two of the three real solutions of the cubic Eq.(\ref{number_photons_cubic_equation}) correspond to high and low oscillation amplitude states with corresponding values of $S_{11}$ from Eq.(\ref{kerr_cavity_S11}), while the third is an unstable, experimentally inaccessible state \cite{Dykman_PRE_2007}.
  
  \begin{figure}[h!]
        \centering
        \includegraphics[width = 0.5\textwidth]{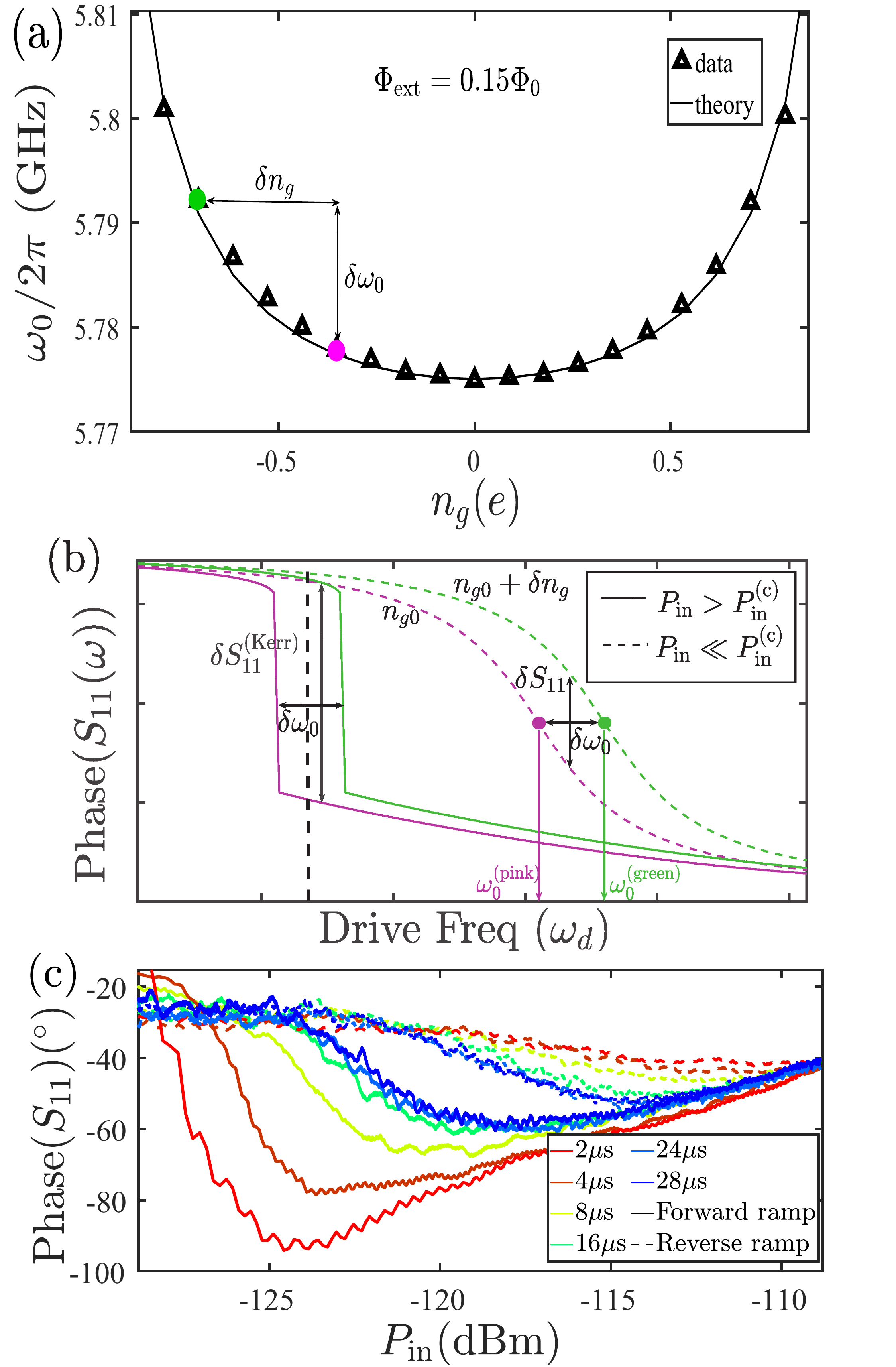}
        \caption{(a) Experimentally measured resonant frequency of the linear cCPT ($P_{\mathrm{in}} \ll P_{\mathrm{in}}^{\mathrm{(c)}}$), $\omega_0$, as a function of the gate charge on the cCPT, $n_g$, at a fixed flux bias $\Phi_{\mathrm{ext}} = 0.15\Phi_0$ (triangles). The line is the theoretically expected resonant frequency for the junction parameters of this device. (b) Simulations illustrating the larger separation in reflected phase, $\delta S_{11}$ ($\delta S_{11}^{\mathrm{(Kerr)}}$) when operating at $P_{\mathrm{in}} > P_{\mathrm{in}}^{\mathrm{(c)}}$ (solid lines) compared to $P_{\mathrm{in}} \ll P_{\mathrm{in}}^{\mathrm{(c)}}$ (dashed lines). (c) Phase of the reflected signal for a forward (solid line) and reverse (dashed line) triangular ramp of the drive amplitude, $P_{\mathrm{in}}$. The input power is ramped between -140 dBm and -109 dBm in increasingly longer times from 2 $\mu$s to 28 $\mu$s from red to blue curves. The cCPT was biased at ($n_g, \Phi_{\mathrm{ext}}$) = (0, 0) and the detuning was $\Delta/2\pi = -9.5$ MHz.}
        \label{fig_2_composite}
    \end{figure}
    
  \par
  The variation of the resonant frequency of the cCPT with the gate, $n_g$, as illustrated in Fig. \ref{fig_2_composite}(a), forms the basis for a sensitive dispersive charge detector. Operating in the single-photon, linear regime, this device was demonstrated to have a charge sensitivity of 14 $\: \mu e/\sqrt{Hz}$ \cite{Brock_PhysRevApplied_2021_ultrasensitive}. Fig. \ref{fig_2_composite}(b) uses Eq.(\ref{kerr_cavity_S11}) to simulate the reflected phase as a function of drive frequency, $\omega_d$, for two gate values separated by $\delta n_g$ corresponding to a resonant frequency shift $\delta \omega_0$. The $S_{11}$ for the two gate values are denoted by pink and green curves respectively, both in the linear ($P_{\mathrm{in}} \ll P_{\mathrm{in}}^{\mathrm{(c)}}$), single photon regime (dashed lines); and with a drive power $P_{\mathrm{in}}> P_{\mathrm{in}}^{\mathrm{(c)}}$ (solid lines). At a given $\omega_d$, $n_g$ can hence be inferred from the measured $S_{11}$. For $P_{\mathrm{in}} > P_{\mathrm{in}}^{\mathrm{(c)}}$, these simulations describe what we would observe in the absence of spontaneous transitions between the high and the low oscillation amplitude states in the bistable region. In the absence of these transitions, while ramping the drive detuning from a large blue-detuned value (with respect to linear resonance, $\omega_0$, $\Delta > 0$), to a red-detuned value ($\Delta < 0$), we expect to stay in the high oscillation amplitude state until we reach the bifurcation detuning further from $\omega_0$ for the green curve in Fig. \ref{fig_1_composite}(c-e). We refer to this as the lower bifurcation point, while referring to the bifurcation detuning closer to $\omega_0$ as the upper bifurcation point. Upon crossing the lower bifurcation point, an abrupt jump from the high to the low oscillation amplitude state is expected, with a corresponding large change in $S_{11}$, as illustrated in Fig. \ref{fig_2_composite}(b). For an appropriate drive frequency denoted by the dashed black line, the same separation in gate charge, $\delta n_g$, produces a larger difference in the reflected phase between the pink and green curves, $\delta S_{11}^{\mathrm{(Kerr)}}$, than the $\delta S_{11}$ while operating in the linear regime. Conversely, $\delta S_{11}^{\mathrm{(Kerr)}}$ continues to remain large as the green and pink curves are brought together by reducing $\delta n_g$, whereas $\delta S_{11}$ undergoes substantial reduction while doing so. The sensitivity of the charge detector is the smallest $\delta n_g$ that produces a $\delta S_{11}$ which can be detected with a signal-to-noise-ratio (SNR) of 1 \cite{Brock_PhysRevApplied_2021_ultrasensitive}. Given that the noise in the measurement is limited by the amplifier chain in the experimental setup \cite{Brock_PhysRevApplied_2021_ultrasensitive}, the larger $S_{11}^{\mathrm{(Kerr)}}$ for smaller $\delta n_g$ promises a lower, much improved charge sensitivity for the device operating in this regime.
    
    \section{Experiments}
    \label{experiment_section}
    \par
    In this section, we first describe the results of a triangular input power ramp in order to understand the extent of the bistable region with respect to the cCPT drive parameters at a given bias point. We then outline the protocol we use in order to perform charge sensing based on the bifurcation described above. Contrary to the sharp jump in $S_{11}^{(\mathrm{kerr})}$ at a precise value of the detuning described in Sec. \ref{theory_section}, we see a non-zero probability of obtaining a value on either end of the step illustrated in Fig. \ref{fig_2_composite}(b) for a range of detunings. We discuss the results of this protocol for a range of cCPT bias points and drive parameters. From this, we glean the optimal conditions for charge sensing and finally perform an optimized single-shot measurement. 
    
    \par
    In order to study the extent of the bistability, we bias the cCPT at ($n_g, \Phi_{\mathrm{ext}}$) = (0, 0), and drive it at a fixed detuning $\Delta/2\pi = -9.5$ MHz with a triangular amplitude ramp in the forward and the reverse direction to check for hysteresis. This is the bias point at which we expect minimum fluctuation in the resonant frequency of the cCPT due to charge and flux noise \cite{Brock_PhysRevApplied_2021, Kanhirathingal_PhysRevAppl_2022}. We perform a heterodyne measurement to obtain the phase of the reflected signal over the course of the ramp. The RF circuitry used in the experiments described here is detailed in Appendix \ref{appendix_experiment_circuit}. Fig. \ref{fig_2_composite}(c) plots the observed hysteresis in the phase of $S_{11}$ for different ramp rates, each averaged over 5000 repetitions of the ramp. For fast ramps, we see that we obtain a value for the reflected phase corresponding to the low oscillation amplitude state for the forward ramp, and a value that corresponds to the high oscillation amplitude state during the reverse ramp. However, as the ramp time is increased, we observe that the spacing between the observed phase during the forward and the reverse ramps reduces, and for this cCPT bias point, saturates to the values represented by the blue curves, corresponding to ramp times of $\sim 25 \: \mu$s. This is because, when given enough time to do so, the oscillator system undergoes fluctuation-induced spontaneous switching between the high and low oscillation amplitude states over the course of a ramp. This yields a weighted average value for the phase at each $P_{\mathrm{in}}$ value over 5000 repetitions of the pulse sequence. The weights depend on the average lifetimes of the high and the low oscillation amplitude states at the chosen cCPT bias and the drive parameters. We see less variation in the shape of the forward and reverse ramp curves for the larger ramp times in Fig. \ref{fig_2_composite}(c). This provides a rough estimate of $\sim 25 \: \mu$s for the average lifetimes of these bistable states. This is a sign of spontaneous transitions between the high and low oscillation amplitude states for a range of input drive strengths, and will be detrimental to the charge sensing scheme described above which counts on the sharp jump from one oscillation amplitude state to the other at precisely a bifurcation point. A similar reduction in the area enclosed between the curves corresponding to the forward and reverse ramps for longer ramps was recently observed for a nonlinear semiconductor microcavity \cite{RodriguezPRL2017}. 
    
    \begin{figure}[h!]
        \centering
        \includegraphics[width = 0.5\textwidth]{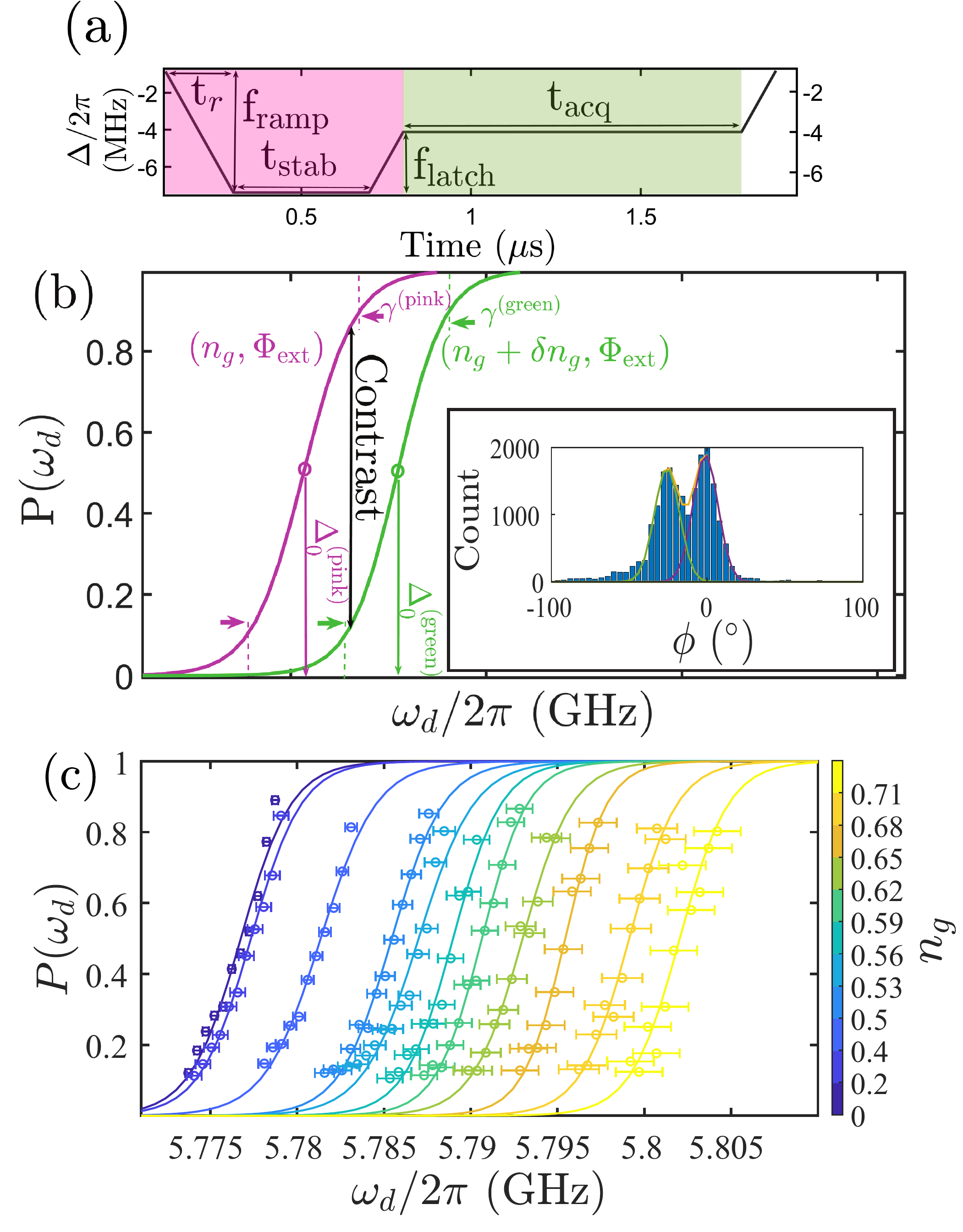}
        \caption{(a) Charge sensing protocol. Detuning of the input pulse tone used to initiate and readout the oscillation state in the charge sensing experiment described in the text with representative values for the durations and detunings of the different sections. The pink area depicts the initialization segment to initialize the oscillator in the high oscillation amplitude state. The phase is measured and averaged during the green segment, for a time $t_{\mathrm{acq}}$. We wait a time $t_{\mathrm{down}} = 5 \: \mu$s between consecutive pulses and set $f_{\mathrm{latch}} = 0$. (b) Schematic illustrating sigmoid  S-curves for two different cCPT gate biases illustrated in pink and green, with the black arrow denoting the maximum contrast between the two. The inset shows a representative histogram of the reflected phase, $\phi$, upon running the above pulse sequence $N_{\mathrm{tot}} = 20,000$ times. The two Gaussian distributions correspond to the oscillator being in the high (left) and low (right) oscillation amplitude states respectively, with the solid lines representing a double Gaussian fit. (c) Obtained S-curves ($\Circle$'s) for different $n_g$ values at $\Phi_{\mathrm{ext}} = 0.06\Phi_0$ and an input drive power $P_{\mathrm{in}} = -128$ dBm. The averaging time, $t_{\mathrm{acq}} = 3 \: \mu$s. The solid lines are sigmoid fits to Eq. (\ref{sigmoid_equation}). The horizontal error bars represent the standard deviation of the resonant frequency fluctuations due to charge and flux noise at the cCPT DC bias point \cite{Brock_PhysRevApplied_2021, Brock_PhysRevApp_2020}.}
        \label{fig_3_composite}
    \end{figure}
    
    \par
    While performing the charge sensing measurement, we choose an input drive strength which gives rise to a region of bistability ($P_{\mathrm{in}} > P_{\mathrm{in}}^{\mathrm{(c)}}$) for the chosen cCPT DC bias $(n_g, \Phi_{\mathrm{ext}})$ with a corresponding $K<0$. In order to deterministically initialize the oscillator in the high oscillation amplitude state, we perform a linear ramp on the detuning of the drive tone from a blue- to a red-detuning as illustrated in Fig. \ref{fig_3_composite}(a). More details on the initialization section (shaded pink) of this protocol are provided in Appendix \ref{appendix_charge_sensing_protocol}. Once initialized, we measure and average the phase of the reflected signal for a time $t_{\mathrm{acq}}$. Performing this measurement $N_{\mathrm{tot}} = 20,000$ times, we obtain a double Gaussian histogram as illustrated in the inset of Fig. \ref{fig_3_composite}(b), and extract the probability of the high oscillation amplitude state, $P(\omega_d)$, as the ratio of the area of the left Gaussian to the total area of the histogram. We perform this measurement for different detunings at the end of the initialization step of the pulse in Fig. \ref{fig_3_composite}(a), and plot the obtained probability of being in the high oscillation amplitude state for each detuning, obtaining the S-curves in Fig. \ref{fig_3_composite}(c). We fit sigmoids of the form 
    \begin{equation}
    P(\omega_d) = \frac{1}{1 + \exp{-\frac{4.3944(\omega_d - \Delta_0)}{\gamma}}},
    \label{sigmoid_equation}
    \end{equation} 
    where $\Delta_0$ is the center of the sigmoid, and the numerical factor in the exponential ensures $\gamma$ is its width between $P(\omega_d) = 0.1$ and $P(\omega_d) = 0.9$.
    
    \par
    As described earlier in Sec. \ref{theory_section}, we ideally expect an abrupt step in $P(\omega_d)$ from $1 \rightarrow 0$ at the lower bifurcation point for our ramp protocol in Fig. \ref{fig_3_composite}(a). However, from Fig. \ref{fig_3_composite}(c), we clearly do not see an abrupt step in $P(\omega_d)$ at just the bifurcation point, but a gradual change in its value across a range of detunings, whose behavior for different cCPT bias points and drive parameters we will now study. 
    
    \par
    For systems where the ratio $\frac{\vert K \vert}{\kappa_{\mathrm{tot}}} < 1$ \cite{AndersenPhyRevAppl2020}, close to a bifurcation, the switching between these two metastable oscillation amplitude states is described by a quantum activation model which predicts fluctuation-induced escape over a metapotential barrier \cite{DykmanPhysica1980, DykmanZhEkspTeorFiz1988}. This has been demonstrated to accurately model the switching between these states in nanomechanical systems \cite{AldridgePRL2005, StambaughPRB2006}, Josephson bifurcation amplifiers \cite{Vijay_Sci_rev_instrum_2009}, and in Josephson junction array devices \cite{Tancredi_APL_2013, Muppalla_PRB_2018}. For systems with Kerr strengths comparable to the cavity linewidth, a quantum calculation is required to accurately model this switching \cite{AndersenPhyRevAppl2020}. We discuss some of the possible sources of these fluctuations in Sec. \ref{discussion_section}.
    
    \par
    From a charge sensing point of view, we want the S-curves for two cCPT gate biases separated by a given $\delta n_g$ to have a large separation between their centers, $\Delta_0$, while the widths of these sigmoids, $\gamma$, should remain small. Additionally, in order to perform single-shot measurements separating the oscillation state using a threshold phase value at the middle of the two Gaussian peaks in the inset in Fig. \ref{fig_3_composite}(b), we need to minimize the overlap between the Gaussians. 
    
     \begin{figure}[h!]
        \centering
        \includegraphics[width = 0.5\textwidth]{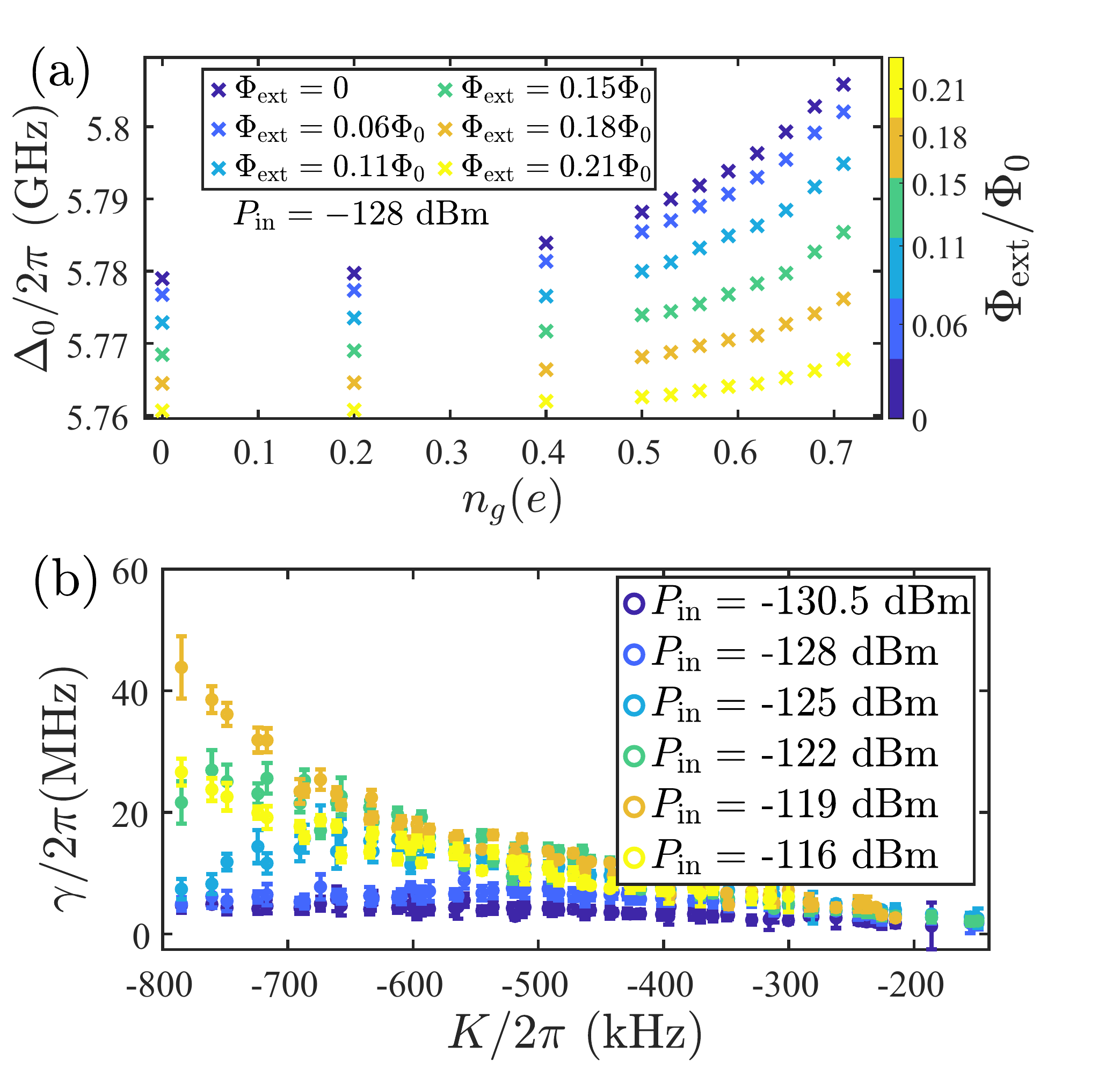}
        \caption{(a) S-curve centers, $\Delta_0$, vs gate charge, $n_g$, for different different flux biases, $\Phi_{\mathrm{ext}}$, for $P_{\mathrm{in}} = -128$ dBm. The error bars are smaller than the markers. (b) S-curve widths, $\gamma$, vs $K$ at the cCPT DC bias points in (a), for different drive strengths. The error bars are the 95$\%$ confidence intervals to the sigmoid fits.}
        \label{fig_4_composite}
    \end{figure}
    
    Fig. \ref{fig_4_composite}(a) shows the variation of the centers of the sigmoid, $\Delta_0$, vs cCPT gate bias, $n_g$, for different cCPT flux biases, $\Phi_{\mathrm{ext}}$. For each flux bias, the largest separations between the centers of two S-curves are observed for large gate biases. Given that we work close to $n_g = 0.71$, the largest separation is for flux values close to $\Phi_{\mathrm{ext}} = 0$. This is related to both the large variation in the ground state energy of the cCPT at these DC bias points, and consequently the linear resonance frequency, $\omega_0$, as in Fig. \ref{fig_2_composite}(a) \cite{Brock_PhysRevApplied_2021}, as well as the variation of the metapotential landscape in the bistable region that limits the extent of switching between the two oscillation amplitude states at a given set of drive parameters. The separation between the $\Delta_0$ for two distinct cCPT bias points is also found to be largest at low input powers, $P_{\mathrm{in}}$ (data not shown). Fig. \ref{fig_4_composite}(b) shows the variation of the width of the sigmoids, $\gamma$, plotted  against the theoretically computed value of the Kerr coefficient, $K$, from Fig. \ref{fig_1_composite}(b) at different cCPT bias points $(n_g, \Phi_{\mathrm{ext}})$, for different $P_{\mathrm{in}}$. The cCPT DC bias points we are interested in based on the separation of the centers of the sigmoids in Fig. \ref{fig_4_composite}(a) correspond to $K/2\pi = -600$ to -800 kHz, and we see $\gamma$ is much smaller for lower $P_{\mathrm{in}}$ at these bias points. This is related both to the reduction in width of the bistable region with decreasing $P_{\mathrm{in}}$, as seen from Eq.(\ref{number_photons_cubic_equation}), and to the reduction in barrier height of the metapotential with increasing $P_{\mathrm{in}}$ \cite{AndersenPhyRevAppl2020, DykmanPRE2007}. This reduced barrier height enables switching between the two oscillation amplitude states within experimental timescales over a wider range of detunings.
    
    \begin{figure}[h!]
        \centering
        \includegraphics[width = 0.5\textwidth]{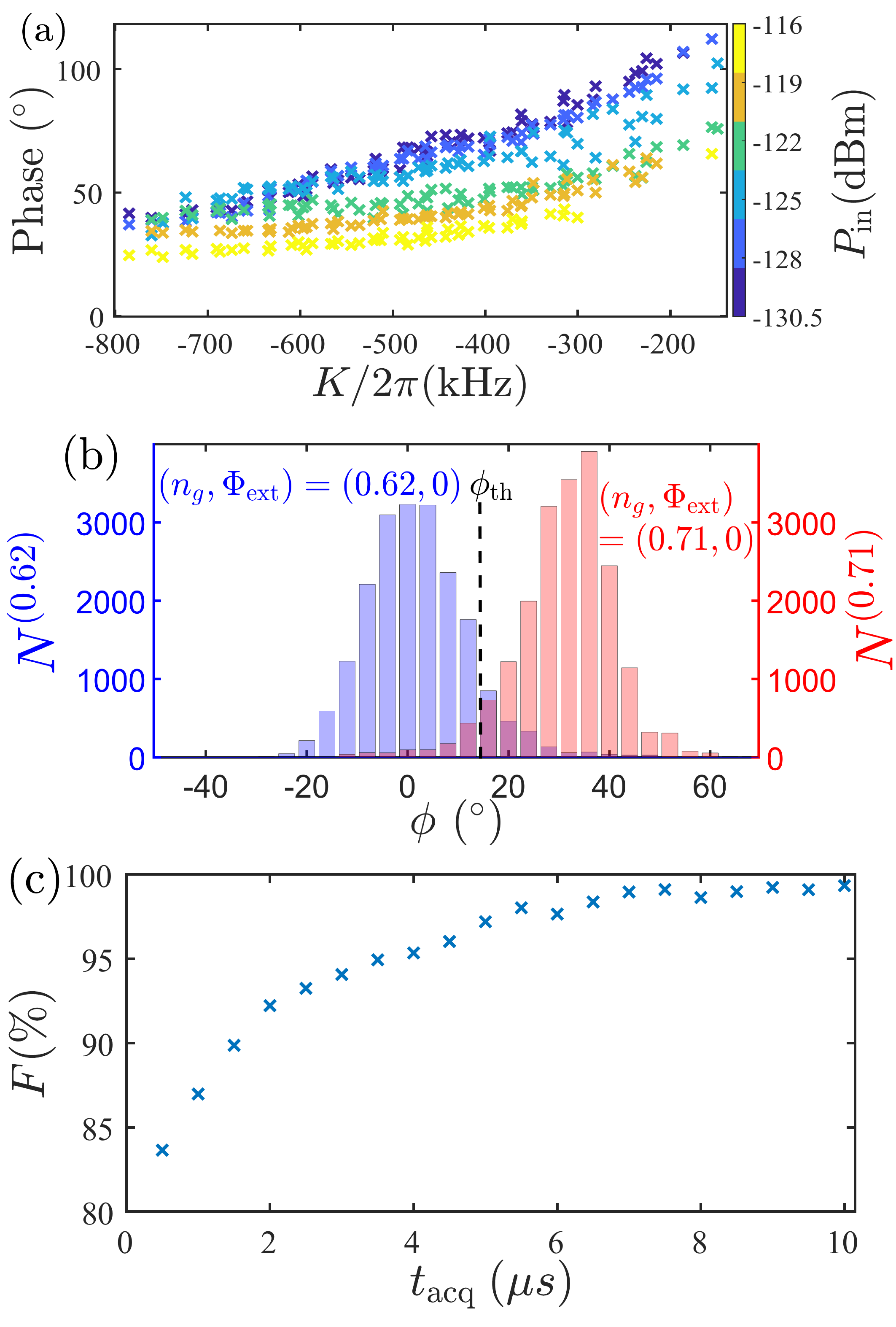}
        \caption{(a) Measured separation between peaks of Gaussians as in inset of Fig. \ref{fig_3_composite}(b) vs $K$ for different $P_{\mathrm{in}}$. (b) Histogram count $N^{(0.62)}$ ($N^{(0.71)}$) of the phase of the reflected signal for optimal charge sensing. The data is for $N_{\mathrm{tot}} = 20,000$ trials each at $(n_g, \Phi_{\mathrm{ext}}) = (0.62,0)$ (blue) and $(n_g, \Phi_{\mathrm{ext}}) = (0.71,0)$ (red) respectively, while driving the cCPT with an input tone at $\omega_d/2\pi$ = 5.8013 GHz with $P_{\mathrm{in}} = -128$ dBm. The dashed line denotes the threshold phase, $\phi_{\mathrm{th}}$, used to discriminate the charge state in a single-shot. (c) Measurement fidelity as a function of averaging time $t_{\mathrm{acq}}$ for the drive conditions in (b).}
        \label{fig_5_composite}
    \end{figure}
    
    \par
    The other consideration in demonstrating single-shot readout is the resolution of the two Gaussians in the inset of Fig. \ref{fig_3_composite}(b). Fig. \ref{fig_5_composite}(a) shows the separation between the centers of the two Gaussians for all the cCPT DC bias points in Fig. \ref{fig_4_composite}(a) for a range of $P_{\mathrm{in}}$. The smallest Kerr strengths, $\vert K \vert$, and the lowest drive strengths, $P_{\mathrm{in}}$, give us maximum separation. We understand this as follows. The detuning corresponding to the maximum oscillator response is either negative (spring softening) or positive (spring hardening), based on $\sgn({K})$. The degree of softening of the oscillation response curve (Fig. \ref{fig_1_composite}(c)) and hence the reflection coefficient, $S_{11}$, illustrated in Fig. \ref{fig_1_composite}(d-e) depends on $\vert K \vert$. The low oscillation amplitude response is close to zero at all detunings in the bistable region, with the corresponding $S_{11}$ close to 1. Meanwhile, the high oscillation amplitude increases from close to zero at the upper bifurcation point to a maximum value at the lower bifurcation point, with a slope inversely proportional to $\vert K \vert$, with similar behavior for $S_{11}$. At detunings close to the upper bifurcation point around which we see non-zero probability for both high and low oscillation amplitude states within experimental lifetimes, the high amplitude oscillation response and the corresponding Phase$(S_{11})$ at a given detuning assume a finite non-zero value whose magnitude depends inversely on $\vert K \vert$, as seen in Fig. \ref{fig_5_composite}(a), such that the difference in Phase($S_{11})$ between the high and the zero phase low amplitude state also depends inversely on $\vert K \vert$. 
    
    \par 
    To understand the variation with $P_{\mathrm{in}}$, we compare the blue and the green curves in Figs. \ref{fig_1_composite}(c-e), which are both for the same $K$. We see that the slope of the amplitude response of both bistable states is nearly independent of $P_{\mathrm{in}}$, but close to the upper bifurcation point, the corresponding $S_{11}$ yields smaller separation between the high and low oscillation amplitude states for higher $P_{\mathrm{in}}$. Fig. \ref{fig_5_composite}(a) hence suggests that we work at low $\vert K \vert$ and at low $P_{\mathrm{in}}$ in order to observe maximum separation between the reflected phase of the high and the low oscillation amplitude states. While we saw that the latter condition also yields S-curves with the smallest widths (Fig. \ref{fig_4_composite}(b)), Fig. \ref{fig_4_composite}(a) shows that the $\Phi_{\mathrm{ext}}$ corresponding to small $\vert K \vert$ values correspond to poor separation between $\Delta_0$ for two cCPT $n_g$ values, which is contradictory to our goal. The pursuit of low $\vert K \vert$ and low $P_{\mathrm{in}}$ suggests operating the cCPT on the cusp of bistability where a large gain in the dispersive readout is expected at a certain detuning \cite{Tosi_Phys_rev_applied_2019}. However, operation of the device studied here is not possible in that regime as discussed further in Sec. \ref{discussion_section}.
    
    \par
    With these considerations, we obtain a maximum contrast of 96.61$\%$ in the S-curves between when the cCPT is biased at $(n_g, \Phi_{\mathrm{ext}}) = (0.62, 0)$ and at $(0.71, 0)$, and driven at $\omega_d/2\pi = 5.8013$ GHz with $P_{\mathrm{in}} = -128$ dBm. For this drive strength, using an averaging time, $t_{\mathrm{acq}} = 3 \: \mu$s, Fig. \ref{fig_5_composite}(b) shows the obtained histograms with counts $N^{(0.62)}(\phi)$ and $N^{(0.71)}(\phi)$ respectively. The separation between the Gaussian peak centers is 36$^\circ$ for this bias point and $P_{\mathrm{in}}$, and the width of each Gaussian is 12$^\circ$ for this $t_{\mathrm{acq}}$. Using a threshold value $\phi_{\mathrm{th}}$ at the center of the two Gaussian peaks as denoted by the dashed line, we assign a charge state to each histogram data point. Defining the fidelity $F^{(0.62)} = 1 - \frac{1}{N_{\mathrm{tot}}}\sum_{\phi = \phi_{\mathrm{th}}}^{180}N^{(0.62)}(\phi)$ and $F^{(0.71)} = 1 - \frac{1}{N_{\mathrm{tot}}} \sum_{\phi = -180} ^{\phi_{\mathrm{th}}}N^{(0.71)}(\phi)$, we obtain an average fidelity $F = 94.59 \: \%$. The similarity between the obtained fidelity and the measured maximum contrast which is agnostic to the overlap of the Gaussians caused by the amplifier noise shows that for this $t_{\mathrm{acq}}$, the limiting factor of our measurement is not the signal-to-noise ratio of the amplifier chain, but is the broadening of the S-curves caused by fluctuation-induced switching between the metastable oscillation states. Finally, it is worth noting that using the above DC bias and drive parameters in Eq. (\ref{number_photons_cubic_equation}) along with damping rates $\kappa_{\mathrm{int}}$, $\kappa_{\mathrm{ext}}$ extracted as described in \cite{Brock_PhysRevApplied_2021, Brock_PhysRevApp_2020}, we find that the intracavity photon number, $n = 8.1$ at $n_g = 0.62$, and $n = 20.94$ at $n_g = 0.71$ respectively in the high oscillation amplitude state. At $n_g = 0.71$, for the optimal drive tone, the oscillator resides predominantly in the low oscillation amplitude state with an intracavity occupation on the order of 0.2 photons. These are orders of magnitude fewer photons than used by devices such as the rf-SETs \cite{Schoelkopf_Science_1998, Andresen_JAP_2008}.
    
    \section{Discussion}
    \label{discussion_section}
    \par
    The cCPT operating in the Kerr bistable regime is thus sensitive to changes in its electrostatic environment that produce a shift of $\delta n_g = 0.09e$ and we have demonstrated real-time single-shot high fidelity detection of this charge difference in $3 \: \mu$s. This corresponds to a charge sensitivity per unit bandwidth  $=\delta n_g \sqrt{t_{\mathrm{acq}}} = 155.89 \: \mu e/\sqrt{\mathrm{Hz}}$. The bandwidth of this electrometer is set by $\kappa_{\mathrm{tot}}/2\pi \approx 1.5$ MHz. This readout is performed with only a few tens of intracavity photons, which is several orders of magnitude smaller than in other state-of-the-art devices \cite{Schoelkopf_Science_1998, Bell_PRB_2012}. An application of the cCPT as a charge sensor is to detect the state of a quantum-dot spin qubit using spin-to-charge conversion \cite{Morello_Nature_2010, He_nature_2019, keith_PRX_2019}. The backaction by the charge sensor on the system being measured is proportional to the number of intracavity photons \cite{Clerk_RevModPhys_2010}, making such low cavity number operation desirable \cite{Aassime_PRL_2001}. Using techniques such as defining the SET in the Si substrate \cite{Morello_PRB_2009, Fuhrer_nanoletters_2009}, and extending the cCPT island \cite{Lu_Nature_2003} in order to increase the $\delta n_g$ induced on the cCPT island in the event of a spin tunneling out of a quantum dot, we could work at larger $P_{\mathrm{in}}$ for the same $\Phi_{\mathrm{ext}}$ and corresponding $K$, while still achieving sufficient contrast and comparable fidelity for much smaller $t_{\mathrm{acq}}$. For the more relaxed $\delta n_g$ requirement, low power operation with smaller $t_{\mathrm{acq}}$ without compromising fidelity would be possible at other $\Phi_{\mathrm{ext}}$ corresponding to smaller $\vert K \vert$ and larger phase separation between the high and the low oscillation amplitude states as in Fig. \ref{fig_5_composite}(a), while still retaining a large contrast value. 
    
    \par
    The major limitation to this mode of operation of the cCPT as a charge sensor are the spontaneous fluctuation-induced transitions between the high and low oscillation amplitude states in the bistable regime. The metapotential landscape governing these transitions depends on $P_{\mathrm{in}}$ and $K$ \cite{DykmanPRE2007, AndersenPhyRevAppl2020}. The $\vert K \vert/\kappa_{\mathrm{tot}}$ range of the cCPT lies in the interesting `mesoscopic' region where quantum effects begin to become important \cite{AndersenPhyRevAppl2020}. Mapping the metapotential and corresponding switching rates between the high and low oscillation amplitude state for such a device could guide understanding of single-photon Kerr devices \cite{Yamaji_PRA_2022} which have been proposed as single-photon sources \cite{Gullans_PRL_2013}, to generate ultra-fast pulses \cite{Fisher_APL_1969} and to be used to implement quantum non-demolition measurements \cite{Grangier_Nature_1998}. 

    \par
    For a given metapotential, the intensity of the fluctuations present in the system is the other factor that affects the switching rates and hence the width of the S-curves, $\gamma$. Given that thermal activation is unlikely since $\hbar \omega_d > k_B T$, one commonly studied source of fluctuations is the dephasing of the oscillator caused by the modulation of the resonant frequency \cite{Dykman_PRE_2007}, or equivalently, of the detuning of the drive. The phase noise of the signal generator is typically $1/f$ in nature and is quite small in the frequency range relevant for escape dynamics such as observed in Fig. \ref{fig_2_composite}(c). The resonant frequency fluctuations for systems such as the cCPT has been studied in detail \cite{Brock_PhysRevApp_2020}, and characterized \cite{Brock_PhysRevApplied_2021}. The resonant frequency fluctuations due to charge noise arising from fluctuating two-level systems close to the cCPT island, and magnetic flux noise arising due to unpaired surface spins are both $1/f$ in character, and should not be relevant to the switching dynamics either. However, the frequency modulation due to white photon shot noise \cite{Clerk_RevModPhys_2010} induced Kerr shift considered in Ref. \cite{Brock_PhysRevApp_2020} would explain the increase in the width of the S-curves at larger $\vert K \vert$ as in Fig. \ref{fig_4_composite}(b), working in tandem with the reduced barrier metapotential barrier at larger $\vert K \vert$. A careful study showing a direct correlation between the switching rates of the cCPT and $P_{\mathrm{in}}$ would confirm this hypothesis, since the frequency independent power spectral density of the photon shot noise depends on the average cavity occupation, $n$.  
    
    \par
    Another avenue to increase the sensitivity of the device would be to reduce the quasiparticle poisoning (QP) in the device. We observe substantial switching out of the even to the odd manifold of the CPT ($n_g \rightarrow 1 - n_g$) \cite{Aumentado_PRL_2004} for $\vert n_g \vert > 0.71$ \cite{Brock_PhysRevApplied_2021}. As illustrated in Fig. \ref{fig_2_composite}(a), the cCPT resonance frequency, $\omega_0$, is most sensitive to $n_g$ close to $\vert n_g \vert = 1$, and employing techniques such as effective  shielding from Cooper-pair-breaking, quasiparticle generating radiation \cite{Barends_APL_2011, Corcoles_APL_2011} could greatly enhance the performance of the cCPT.

    \par
    Still using this inherent Kerr nonlinearity, but by driving the cCPT with a $P_{\mathrm{in}} \sim P_{\mathrm{in}}^{\mathrm{(c)}}$ close to but before the onset of bistability where $\frac{dS_{11}}{d\Delta} \rightarrow \infty$ for some $\Delta$, we should be able to realize a large gain in the charge sensitivity \cite{Tosi_Phys_rev_applied_2019}. The presence of gate and flux noise induced resonance frequency fluctuations \cite{Brock_PhysRevApplied_2021, Brock_PhysRevApp_2020} make it hard to operate at the precise $\Delta$ where this enhancement is expected, but using the resonance frequency stabilizing feedback scheme demonstrated in \cite{Kanhirathingal_PhysRevAppl_2022} should enable such operation. Enhanced cooling of a nanomechanical resonator coupled to a nonlinear cavity operating in this regime has been shown \cite{Zoepfl_Arxiv_2022}.
    
    \par
    The cCPT Hamiltonian also has other nonlinear terms such as those of a degenerate parametric amplifier which can be driven into resonance using an appropriate flux pump at close to $2\omega_0$. The amplitude of the parametric oscillations induced \cite{Wilson_PRL_2010, Wustmann_PRB_2013, Krantz_NJP_2013} depends on the gate bias of the cCPT \cite{Thyagarajan_thesis_2021}, and can be used as a charge sensor, similar to the qubit state detector operating on this principle \cite{Krantz_Nature_2016}. 
    
    
    \begin{acknowledgments}
    We thank Ethan Williams, Hui Wang, Archana Kamal, Chandrasekhar Ramanathan, William Braasch and Hory Mohammadpour for helpful discussions and useful feedback on the manuscript. This work was supported by the NSF under Grants No. DMR-1807785 (S. K., B. T., B. L. B., and A. R) and DMR-1507383 (M. P. B.), and by a Google research award (S. K.).
    \end{acknowledgments}
    
    \appendix
    \section{Experimental setup}
    \label{appendix_experiment_circuit}
    
     \begin{figure}[h!]
        \centering
        \includegraphics[width = 0.5\textwidth]{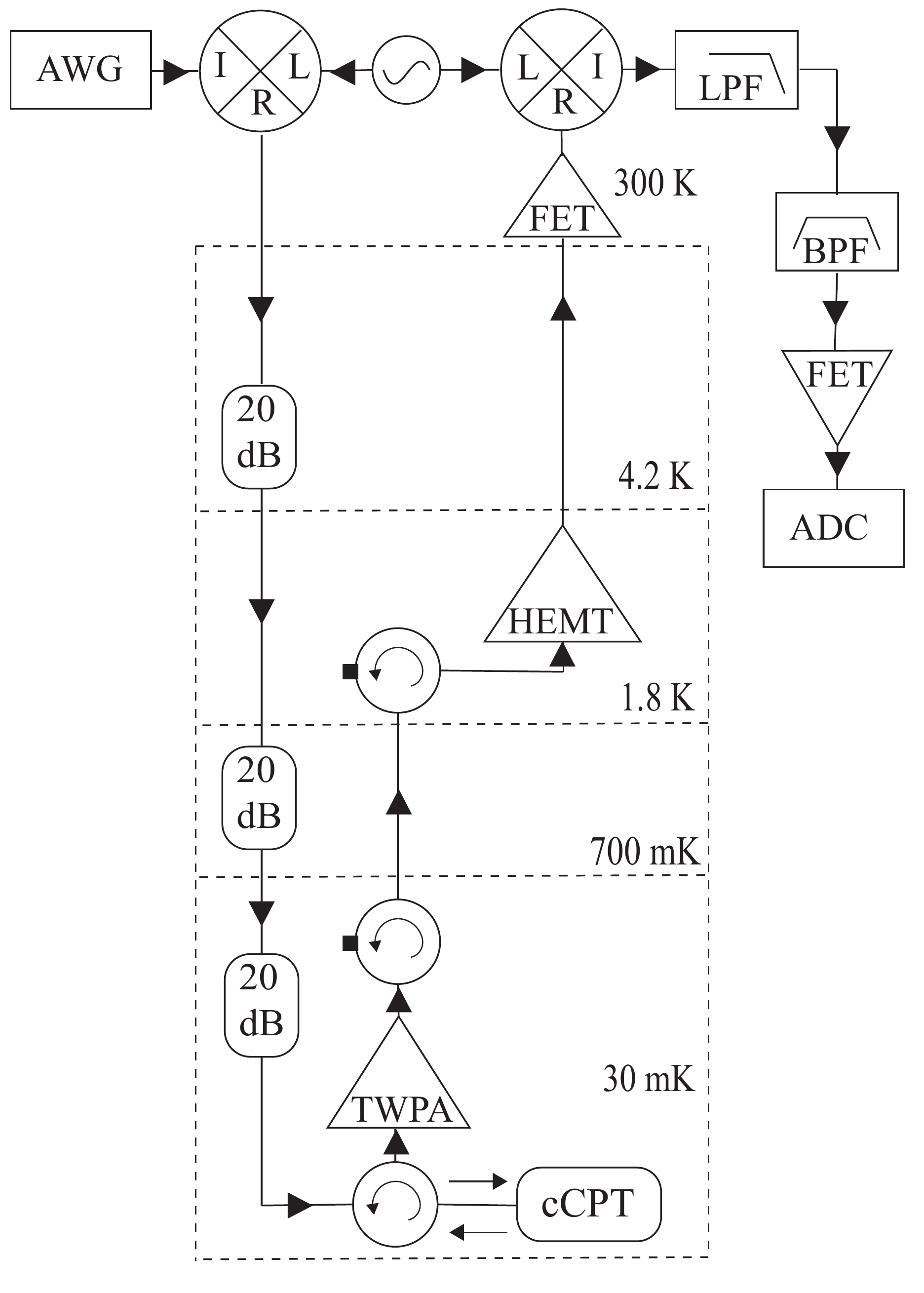}
        \caption{Microwave circuitry used in Sec. \ref{experiment_section}.}
        \label{fig_circuit_diagram}
    \end{figure}

    \par
    Fig. \ref{fig_circuit_diagram} shows the rf circuitry used in the experiments described in this work. The input tone from a Keysight N5183B signal generator is mixed with an intermediate frequency tone from a Tektronix arbitrary waveform generator whose amplitude envelope can be ramped, or whose frequency can be chirped for the charge sensing protocol (see Fig. \ref{fig_3_composite}(a)). This signal passes through various stages of attenuation in the dilution refrigerator before driving the cCPT, which is mounted inside a magnetic shield at the mixing chamber stage of the refrigerator. The reflected signal goes through a circulator to a traveling wave parametric amplifier (TWPA) \cite{Macklin_science_2015} which serves as the first stage amplifier. The signal is then amplified by a Low Noise Factory LNF LNC4$\textunderscore$8C high electron mobility transistor (HEMT) and a room temperature low noise field effect transistor (FET). This signal is then mixed down to an intermediate frequency of 21 MHz, filtered, digitally sampled, and demodulated to extract the phase. 

    \par
    The TWPA has an average gain of 18 dB over the operating bandwidth of the cCPT, ensuring that the added noise of the amplifier chain is dominated by the noise added by the TWPA. The added noise density referred to the input of the amplifier chain is separately measured to be $\sim$4.67 photons/Hz (noise temperature of 1.28 K), close to the quantum limit of 1 photon/Hz \cite{Clerk_RevModPhys_2010} for the phase insensitive TWPA. 

    \section{Charge sensing protocol initialization}
    \label{appendix_charge_sensing_protocol}
    Here, we elaborate upon the initialization section of the charge sensing protocol illustrated in Fig. \ref{fig_3_composite}(a). In order to initialize the oscillator in the high oscillation amplitude state, we start from a detuning in the monostable region on the positive detuning side in Fig. \ref{fig_1_composite}(c-e), and ramp the detuning by $f_{\mathrm{ramp}}/2\pi = - 41$ MHz in a time $t_r = 530$ ns. The final detuning is close to a bifurcation edge, denoted by the black dashed line in Fig. \ref{fig_2_composite}(b). The oscillator is driven with this constant tone for a time $t_{\mathrm{stab}} = 4.9 \: \mu$s, during which time fluctuations could cause a transition to the low oscillation amplitude state. These values for $t_r$ and $t_{\mathrm{stab}}$ were settled upon after performing QuTiP \cite{Johansson_qutip_2013} simulations using a master equation solver for the exact input tone in Fig. \ref{fig_3_composite}(a), and seeing the system through a transient evolution period to the steady state. This value of $t_{\mathrm{stab}}$ is also close to the nominal value of 5/$(\kappa_{\mathrm{tot}}/2\pi)$ over which transients of oscillating systems are expected to decay, even in the region where switching between high and low oscillation amplitude states is observed, where the oscillator dynamics are considerably slowed \cite{AndersenPhyRevAppl2020}. The detuning could then be ramped to a slightly larger blue-detuning, $f_{\mathrm{latch}}$, to reduce the probability of a switching event during the measurement time, hence `latching' the oscillator in the oscillation amplitude state attained at the end of the stabilization time \cite{Vijay_Sci_rev_instrum_2009}. However, unlike the systems studied in Refs. \cite{StambaughPRB2006, Vijay_Sci_rev_instrum_2009}, the low oscillation amplitude state is often not a long-lived state in our system, making the latching a little less likely, and causing the switching statistics to depend on the additional parameters $f_{\mathrm{latch}}$ and $t_{\mathrm{acq}}$. We thus set $f_{\mathrm{latch}} = 0$. 
    
    \bibliographystyle{apsrev4-1-no-url}
    \bibliography{biblio}

\end{document}